# Steep optical-wave group velocity reduction and "storage" of light without electromagnetically induced transparency

M. Kozuma[(1)], D. Akamatsu[(1)], L. Deng[(2)], E.W. Hagley[(2)], and M.G. Payne[(3)]

[(1)] Department of Physics, Tokyo Institute of Technology, 2-12-1 O-okayama, Meguro, Tokyo, Japan 152-8550

[(2)]Electron & Optical Physics Division, NIST, Gaithersburg, MD 20899

[(3)]Department of Physics, Georgia Southern University, Statesboro, GA 30460

We report experimental investigation of optical-pulse group velocity reduction and probe-pulse regeneration using a Raman scheme. The new scheme, which does not rely on electromagnetically induced transparency (EIT), has many advantages over the conventional method that critically relies on the transparency window created by an EIT process. We demonstrate significant reduction of the group velocity, less probe-field loss, reduced probe-pulse distortion, and high probe-pulse regeneration efficiency.

Electromagnetically induced transparency (EIT) [1] is a process where a destructive interference between optically shifted resonances cancel the linear contributions of the dispersion function of a probe pulse that is tuned to the line center. EIT has been used in almost every recent study on the group velocity reduction, atomic spin-wave excitation, and probe-pulse regeneration in resonant medium [2-4]. Strictly speaking, the ideal on-one-photon-resonance EIT is applicable only to a three-level model. In reality, however, the situation is more complicated because no ideal three-level system exists. The effects of nearby hyperfine states often contribute significantly to wave propagation, thereby greatly altering the response of the system to the optical wave. Indeed, it has been shown [5] that in the case of the sodium $D_1$ and $D_2$ lines, the probe-field loss due to nearby hyperfine states generally dwarfs that of the pure three level system, and causes nearly an 80% reduction in the probe field intensity. In addition, it also causes as much as 30% probe-pulse broadening. Even if one chooses an atomic species with a larger hyperfine splitting (e.g., rubidium), there will still be a loss of probe-field intensity, and a distortion of the form of light pulse. Moreover, the stringent requirement on having both the probe and coupling lasers tuned exactly on resonance remains. This latter requirement is one of the major obstacles that prevents the EIT scheme from being applied to solids at room temperature where broad energy-band structures render "tuning to the resonance line center" rather meaningless. To search for a better scheme, we note that EIT is not necessary for achieving significant group velocity reduction. Indeed, significant modification of the dispersion of the medium can be achieved without EIT [6,7].

In this letter, we report experimental results on significant group velocity reduction and probe-pulse regeneration without using EIT [8,9]. The Raman scheme described here has many advantages over the conventional one-photon, on-resonance EIT scheme. The introduction of a non-negligible one-photon detuning opens many possibilities, such as not only a reduction of the probe field loss, but even probe field gain. We will show that by using a conventional vapor cell with properly chosen parameters, this new scheme can greatly reduce probe-field attenuation, and achieve high efficiency probe pulse regeneration. In addition, we will demonstrate smaller pulse-width broadening. To the best of our knowledge, to date none of these features have been reported in the literature.

Our experiment is carried out in a vapor cell that is filled with pure $^{87}$Rb atoms whose partial energy level is depicted in the top panel of Fig.1. The experiment is performed using the D$_1$ line of the $5^2S_{1/2}, F = 2 \rightarrow 5^2P_{1/2}, F = 2$ transition. We use two ground states |0> = |F=2, m$_F$=–2> , |1> = |F=2, m$_F$=0>, and one excited state |2> = |F′=2, m$_F$=–1>. The experimental setup is shown in lower panel of Fig. 1. The temperature of the cell, which is 10 cm long and heated with a non-magnetic wire, is typically in the range of 50-70°C and is stabilized for uniform atomic density. The estimated full width of the Doppler broadened |0>→ |2> transition at this temperature is about 560 MHz. In order to ensure long lifetimes of the atomic Zeeman coherence we use a triple-layer magnetic shield. The Rb cell is also filled with 5 torr of He$^4$ buffer gas so that the Rb atoms stay in the path of the laser beams for several hundreds of microseconds due to elastic collisions with He atoms. Both the coupling ($\sigma^-$) and probe ($\sigma^+$) lasers are derived from a single extended-cavity diode laser. The diameter of laser beam in the vapor cell is 2 mm. We first turn the control laser on the resonance to carry out the optical pumping stage. We then switch the control laser to a predetermined one-photon detuning. At the same time we slightly rotated the polarization of the input light to create a weak $\sigma^+$ probe light pulse (duration $\tau = 7.3$ μs) using a fast Pockels cell made from an 80 mm long LiNbO$_3$ crystal. We chose the power of the control and probe lasers to give single-photon Rabi frequencies of $\Omega_{21} = \Omega_c(\sigma^-) = 2\pi \times 15 MHz$ and $\Omega_{20} = \Omega_p(\sigma^+) = 2\pi \times 1.3 MHz$, respectively. These choices of power and pulse duration ensure that during the presence of the probe pulse there is negligible population transfer; a necessary condition for a simplified perturbative treatment of the system that was used to guide our choice of experimental parameters. Under these conditions, we have $\Delta_{ac}\tau > 1$ ($\Delta_{ac} = \frac{|\Omega_{21}|^2}{\delta_{20}}$ is the ac Stark shift due to the control laser), and $|\delta_{20}| \gg |\Omega_{21}|$. Ideally, one would prefer to have the laser detuned several Doppler line widths while maintaining a large ac Stark shift. However, due to the limited output power of our control laser, we are unable to detune very far from state |2>, while still producing a sizable ac Stark shift. Therefore, the above-described choice of parameters represents a reasonable compromise in view of our experimental constraints.

Before presenting our data, let us briefly state some of the main features that are expected in the Raman scheme for a set of conditions that ensure the validity of an adiabatic treatment. [6,7]. When $\Delta_{ac}\tau \gg 1$, we expect the propagation velocity of the probe pulse to be independent of the detuning, but linearly proportional to the power of the control laser. In addition, we expect a significant reduction of probe field loss. In Figure 2 we show a plot of the probe-pulse intensity profile at the exit of the cell as a function of the time. To better illustrate the advantages of the Raman scheme, we present measurements of both the Raman scheme and the conventional EIT scheme. Three features are immediately apparent. First, a significant reduction of the group velocity is achieved, and the data show that the slow-down effect of the proposed Raman scheme is almost the same as that of the EIT scheme. The second feature exhibited in Fig. 2 is the drastic reduction of the probe-field loss compared to the EIT scheme. In the case of the EIT scheme, the probe field has suffered nearly a 99% loss in intensity, whereas in the Raman case, this loss is only about 58%, yielding a signal nearly forty times as intense as in the EIT case. We have examined the detuning range from 500 MHz to 1.2 GHz. With 1.2 GHz detuning, the Raman scheme preserves nearly 99% of the intensity of the original probe pulse but the group velocity is about a factor of ten faster than with the EIT scheme. The second, and perhaps the most important advantage of the Raman scheme over the EIT scheme is the fact that the large Raman detuning readily allows it to be applied in solid state materials, a territory that prohibits a meaningful application of the EIT scheme because of the broad energy-band structure [8]. The third feature that is noticed from Fig. 2 is the probe pulse line-width broadening. In the Raman scheme, this broadening is about 10% whereas in the EIT scheme, the broadening is about 40%. The smaller pulse broadening of Raman scheme is due to the large Doppler-broadened upper-state line width. It can be shown theoretically, and has been observed experimentally [2], that in the conventional EIT scheme the probe pulse is always broadened during propagation in a true three level system. This occurs because when the ac Stark shift induced by the control laser Rabi frequency is not substantially larger than the upper state lifetime, the atomic dispersion function at the probe frequency cannot be treated in a linear response theory. The consequence is that nonlinear contributions in the dispersion function will lead to pulse broadening regardless of whether the medium is hot or cold. In the Raman case, however, the dynamics are changed by the fact that the one-photon detuning also plays a role in determining the size of the nonlinear contribution, resulting in a less pulse broadening, as is shown in Fig. 2. Experimentally, we also found that in the range studied (500 - 1000 MHz detuning), the group velocity is insensitive to the detuning. Near the end of the range, i.e. near 1.2 GHz detuning, we observed an increase in the group velocity. In addition, we have observed that the group velocity scales linearly with the power of the control laser. These observations are in good agreement with theoretical predictions for the control laser power used in our experiment.

The group velocity achievable with a Raman scheme in a typical three-level system is about the same, and becomes slightly worse than the EIT scheme (typically within a factor of 4) as the detuning increases. This slight difference, however, is a reasonable price to pay for achieving significantly lower loss, less pulse broadening, and interesting pulse-propagation dynamics introduced by the sizable one-photon detuning. These advantages are expected to hold for cold vapors, hot vapors, and even solid medium.

To further demonstrate the significant advantages of the Raman scheme, we have investigated the "storage" of probe photons, and the subsequent recovery of the probe pulse. In Fig. 3, we show the storage and recovery of a probe pulse obtained using the Raman scheme, and compare it with the EIT scheme under the same conditions. The first striking feature of Fig. 3 (left panel) is the significantly lower loss in Raman scheme. In fact, under the same experimental conditions, signals from the EIT scheme are very weak and noisy. In contrast, notice the clean pulse shape obtained using the Raman method. On the right panel, we show a series of probe pulse recoveries with several control laser pulses turned on at different delay times. The cleanly regenerated probe pulse is clear evidence that the Raman scheme is superior to the EIT scheme. Under the same conditions, the EIT method produces a series of pulses (not shown) that are barely above the noise level.

We emphasize that an on-resonance EIT scheme is not necessary for achieving group velocity reduction. The role of EIT process is mainly to reduce one-photon absorption so that the probe pulse can travel with low loss in an otherwise opaque medium, allowing one to measure significant group velocity reduction at low control laser power. In the Raman scheme this is achieved with a non-vanishing one-photon detuning at the expense of higher control-laser power. The advantages of the Raman scheme are very obvious when it is applied to solid medium for group velocity reduction. This is because the broad energy-band structure encountered in solids at room temperature makes the on-one-photon resonance condition required by the conventional EIT scheme very difficult to satisfy. Any part of the energy band that is not strongly overlapped by the coupling laser will not be driven transparent, and hence will collectively contribute as a loss mechanism. However in the case of the Raman scheme, the one-photon detuning can be chosen to be much larger than the energy bandwidth, thereby preserving the probe-pulse intensity while achieving a comparable reduction in the group velocity and yielding a much higher probe-pulse recovery efficiency. This could make our technique very useful in novel optical device designs that may have potential applications to telecommunications.

M.K. and D.A acknowledge the supported from the program of Research and Development of Quantum Communication Technology by the Ministry of Public Management, Home Affairs, Posts and Telecommunications of Japan.

**Figure captions:**

Figure 1.  Upper panel: 3-level EIT scheme (right), and 3-level Raman scheme (left). Energy level designations for Raman scheme: $|0\rangle=|F=2, m_F=-2\rangle$, $|2\rangle=|F'=2, m_F=-1\rangle$, and $|1\rangle=|F=2, m_F=0\rangle$. Lower panel: Experimental setup.

Figure 2.  A plot of the probe pulse as a function of time for an on-resonance (EIT) scheme, and for a Raman scheme and with a detuning of $\delta_{20} = 774$ MHz. The EIT data has been magnified by 83, and the probe profile for the Raman scheme was magnified by 2.4. The estimated group velocity is about $c/10^4$. The closeness of the peak positions indicates that under the driving conditions used the group velocity is insensitive to the detuning, just as the theory predicts.

Figure 3.  Left panel: "storage of light" using a Raman and an EIT scheme. Parameters are similar to that of Fig. 2. Notice the broadened pulse and significantly lower and noisier "probe pulse" regenerated under the EIT scheme.
Right panel: Successive probe pulse regeneration using a series of control laser pulses. The clean probe pulses clearly show that the Raman method is superior to the EIT method under the present conditions.

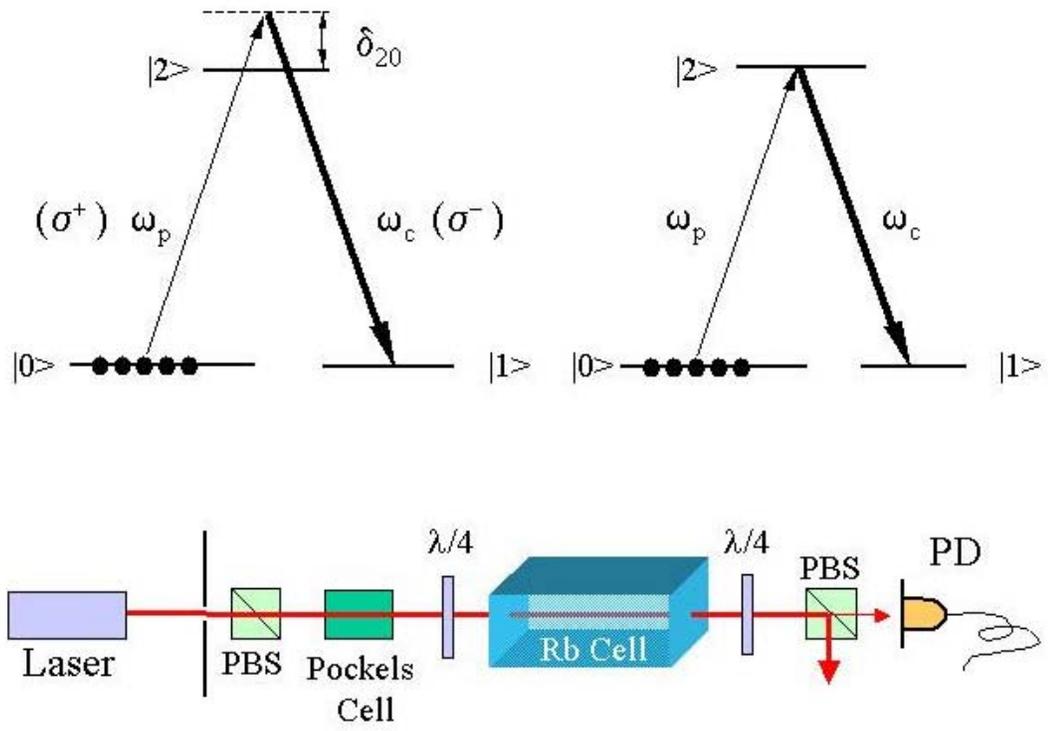

Fig. 1 Kozuma et al.

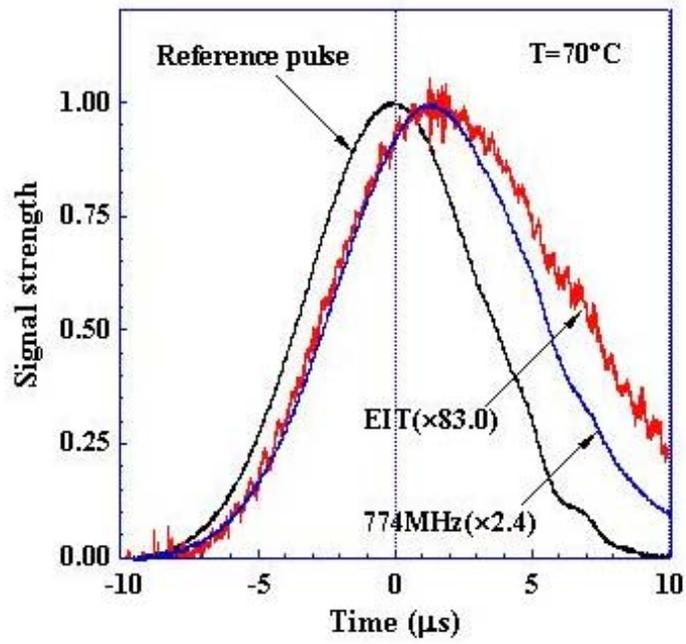

Fig.2 Kozuma et al.

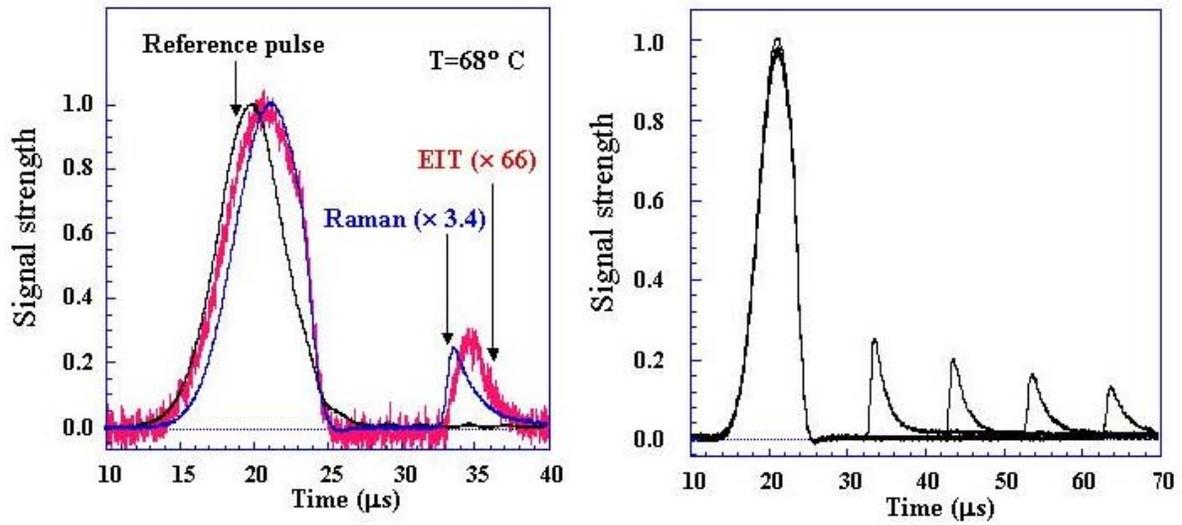

Fig.3 Kozuma, et al